\documentclass[prb,showpacs,twocolumn,superscriptaddress]{revtex4-1}
\usepackage{amssymb}
\usepackage{graphicx}
\usepackage{amsmath,amssymb}

\begin{document}

\title{Strain manipulation of Majorana fermions in graphene}
\author{Zhen-Hua Wang}
\affiliation{Beijing Computational Science Research Center, Beijing 100084,
China}
\author{Eduardo V. Castro}
\affiliation{CeFEMA, Instituto Superior T\'{e}cnico, Universidade de Lisboa,
Av. Rovisco Pais, 1049-001 Lisboa, Portugal}
\author{Hai-Qing Lin}
\affiliation{Beijing Computational Science Research Center, Beijing 100084,
China}

\begin{abstract}
The functionalized graphene with induced superconductivity, Zeeman coupling,
and finite Rashba spin-orbit coupling is proposed to display topological
superconducting phases with Majorana end modes. We obtain the phase diagram
of bulk graphene and nanoribbon by calculating the Chern number, band
structure and wavefunction. The electron doping in graphene, magnetic field
and strain-induced pseudomagnetic field can result in the topological phase
transition. Moreover, it is interested to note that strain has negative
influence on the stability of topological nontrivial phase either uniform or
nonuniform, destroying the existence of Majorana fermion, which provides a
new way to transfer, create and fuse Majorana fermions. Some experimental
schemes are also introduced to tailor functionalized graphene, generating
various devices applied in topological quantum computation.
\end{abstract}

\maketitle

\section{Introduction}

Graphene and related materials can be integrated with desired specific
properties, in particular to the energy devices,\cite
{bonaccorso2015graphene,berman2015macroscale,fan2015m3c,skrypnychuk2015enhanced,georgakilas2012functionalization}
topological quantum computations\cite
{karafyllidis2015quantum,RevModPhys.82.3045,moore2010birth,du2009fractional,bolotin2009observation, stern2013topological,guo2009quantum,kou2013graphene}
and wide spectrum of applications.\cite
{he2010centimeter,premkumar2012graphene,artiles2011graphene,mohanty2008graphene,singh2011graphene,huang2012graphene,ivanovskii2012graphene}
With the development and optimization of production methods, high-volume
liquid-phase exfoliation\cite{hernandez2008high,park2009chemical,wei2012graphene,paton2014scalable} and
chemical vapor deposition,\cite
{raccichini2015role,allen2009honeycomb,ismach2010direct} functionalized
graphene is tailored and exploited for various devices, such as portable and
wearable energy conservation and shortage devices, water splitting,
membranes, quantum walks and computation. Topological properties and the
chiral symmetry inherent in graphene are the main reason why the seemingly
simply honeycomb lattice accommodate such a rich physics,\cite
{hatsugai2014graphene,novoselov2005two,geim2007rise,zhang2005experimental,bhattacharya2011comprehensive}
which includes double Dirac cones in the Brillouin zone,\cite%
{lomer1955valence} anomalously sharp Landau level and quantum Hall effect at
the Dirac point in magnetic field even with ripples, and a host of other
peculiar features.\cite{hatsugai2009bulk,hatsugai2006quantized}

Graphene is one of the most promising nanomaterials because of its unique
combination of superb properties. Experimental advances in doping methods
have allowed the electron density to approach the van Hove singularities
(VHSs) at 25\% hole or electron doping.\cite%
{PhysRevB.92.174503,PhysRevLett.109.197001} The logarithmically diverging
density of states at the VHS can allow nontrivial ordered ground states to
emerge due to strongly enhanced effects of interactions. Meanwhile, strain
is used as a mechanism to change the energy of the VHS, bringing it closer
to the doping level of pristine graphene, and therefore making it
experimentally accessible. Topological insulators is predicted for graphene,
where the weak intrinsic spin-orbit interaction (SOI) can be enhanced by
applying a perpendicular electric field\cite%
{PhysRevLett.105.256805,PhysRevB.79.165442} or a nonuniform magnetic field,%
\cite{PhysRevX.3.011008} tuning the local curvature of the sheet,\cite%
{PhysRevB.74.155426} or doping by 3d or 5d transition metal adatoms.\cite%
{PhysRevLett.104.136803,PhysRevB.85.245441,PhysRevLett.109.266801} When
graphene is deposited on substrates or adsorbed with heavy atoms, the
interaction-induced symmetry breaking can open bulk gaps to support kinds of
topological phases. Recently, one common strategy to generate topological
superconductor is through the coupling of an s-wave superconductor to the
helical half-metallic graphene.\cite{alicea2012new,PhysRevX.5.041042}

The exotic properties of graphene also led to an extensive study focusing on
monolayer, bilayer, nanoribbon, and taking into account features such as
strain and curvature, disorder, etc.\cite%
{PhysRevB.86.235416,PhysRevLett.101.096402} Being able to influence the
motion of charge carriers, strain-induced pseudomagnetic fields in graphene
have been explored as a potential approach to engineering the electronic
states of graphene.\cite{guinea2010energy,PhysRevLett.115.245501}
Nanoribbons also provide numerous interesting issues in fundamental physics,
such as topological insulators.

The functionalized graphene in topological hybrid system has received less
attention, and therefore graphene with induced superconductivity and Zeeman
coupling, and finite Rashba spin-orbit coupling is proposed to display
topological superconducting phases with Majorana end modes. Besides bulk
graphene, nanoribbons are also designed to investigate their topological
phase transition, and the uniform and nonuniform strain induced
pseudomagnetic fields have great effect on the stability of topological
nontrivial phase, which offer new method to transfer and control Majorana
fermions (MF). Nonuniform chemical potential, Zeeman field and strain
provide us enough quantum techniques to control functionalized graphene.

\section{Model}

\begin{figure}[t]
\center{\includegraphics[clip=true,width=\columnwidth]{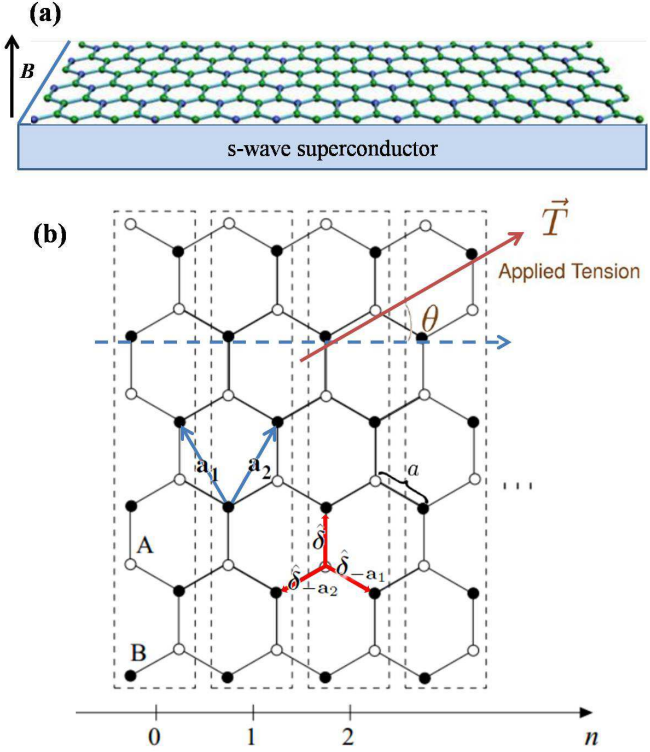}}
\caption{Sketch of zigzag graphene nanoribbon. (a) The zigzag graphene
nanoribbon on an s-wave superconductor, and the magnetic field \textit{B}
applied perpendicular to the surface can induce a topological regime, where
bound states are formed inside the gap. (b) Honeycomb lattice geometry. The
vectors $\hat{\protect\delta},\ \hat{\protect\delta}_{-a_{1}}$ and $\hat{%
\protect\delta}_{-a_{2}}$ connect A sites to their B site neighbors. The
graphene ribbon is stretched or compressed along a prescribed direction, and
$\protect\theta$ is direction of applied tension T. }
\label{fig1}
\end{figure}

In this work only nearest neighbor hopping processes are considered, shown
in Fig. 1, the minimum tight binding model is used:%
\begin{eqnarray}
H_{0} &=&-t\sum\limits_{r,\sigma }a_{r,\sigma }^{+}(b_{r,\sigma
}+b_{r-a_{1},\sigma }+b_{r-a_{2},\sigma })+h.c.  \notag \\
&&-\mu \sum\limits_{r,\sigma }(a_{r,\sigma }^{+}a_{r,\sigma }+b_{r,\sigma
}^{+}b_{r,\sigma }),  \label{EQ1}
\end{eqnarray}%
where $r$ denotes the position on the Bravais lattice, and $t$\ connects the
site $r$ to its neighbors, $\mu $ is the chemical potential, and $%
a_{r,\sigma }$ ($b_{r,\sigma }$) annihilates an electron with spin $\sigma $
on atom A (B) in the unit cell.

The Rashba SOI Hamiltonian is written as
\begin{eqnarray}
H_{R} &=&i\lambda _{R}\sum\limits_{r,\sigma ,\sigma ^{\prime }}a_{r,\sigma
}^{+}[(\sigma \times \hat{\delta})_{z}^{\sigma \sigma ^{\prime }}b_{r,\sigma
^{\prime }}+(\sigma \times \hat{\delta}_{-a_{1}})_{z}^{\sigma \sigma
^{\prime }}b_{r-a_{1},\sigma ^{\prime }}  \notag \\
&&+(\sigma \times \hat{\delta}_{-a_{2}})_{z}^{\sigma \sigma ^{\prime
}}b_{r-a_{2},\sigma ^{\prime }}]+h.c.,  \label{EQ2}
\end{eqnarray}%
where $\lambda _{R}$ is strength of spin-orbit coupling, ($\sigma ,\sigma
^{\prime }$) stand for spin up and down.

The s-wave superconductivity is induced in the system by proximity effect
and is described by a uniform on-site superconducting order parameter $%
\Delta $ whose Hamiltonian can be written as:%
\begin{equation}
H_{SC}=\Delta \sum\limits_{r}a_{r,\uparrow }^{+}a_{r,\downarrow
}^{+}+b_{r,\uparrow }^{+}b_{r,\downarrow }^{+}+h.c..  \label{EQ3}
\end{equation}

Finally, a magnetic field $B$ is introduced, resulting in an out-of-plane
Zeeman potential $V_{Z}=g\mu _{B}|B|/2$, with $g$ the Land\'{e} $g$-factor
and $\mu _{B}$ the Bohr magneton. This Zeeman Hamiltonian is described as
follows:%
\begin{equation}
H_{Z}=V_{Z}\sum\limits_{r,\sigma ,\sigma ^{\prime }}(a_{r,\sigma }^{+}\sigma
_{z}^{\sigma \sigma ^{\prime }}a_{r,\sigma ^{\prime }}+b_{r,\sigma
}^{+}\sigma _{z}^{\sigma \sigma ^{\prime }}b_{r,\sigma ^{\prime }}).
\label{EQ4}
\end{equation}

We choose $a_{1(2)}$\ as basis vectors in direct space

\begin{equation}
\mathbf{a}_{1(2)}=\sqrt{3}a(\mp \frac{1}{2},\frac{\sqrt{3}}{2}),  \label{EQ5}
\end{equation}

and for the vectors connecting NN sites,%
\begin{eqnarray}
\hat{\delta} &=&a(0,1),  \notag \\
\hat{\delta}_{-a_{1}} &=&a(\frac{\sqrt{3}}{2},-\frac{1}{2})  \notag \\
\hat{\delta}_{-a_{2}} &=&a(-\frac{\sqrt{3}}{2},-\frac{1}{2}).  \label{EQ6}
\end{eqnarray}

The basis vectors in reciprocal space may be chosen as,

\begin{equation}
\mathbf{b}_{1(2)}=\frac{4\pi }{3a}(\mp \frac{\sqrt{3}}{2},\frac{1}{2}),
\label{EQ7}
\end{equation}%
and the corners of the hexagonal Brillouin zone, where the gap closes, are
given by%
\begin{equation}
\pm \mathbf{K}=\pm \frac{\mathbf{b}_{2}-\mathbf{b}_{1}}{3}=\frac{4\pi }{3a}%
\hat{e}_{x}\ .  \label{EQ8}
\end{equation}

We are\ also interested in uniform planar tension situations, where the
graphene sheet is stretched or compressed uniformly along a prescribed
direction. The fixed Cartesian system is chosen in a way that $Ox$ always
coincides with the zigzag direction of the lattice. The strain tensor in the
lattice coordinate system reads:\cite{PhysRevB.80.045401}%
\begin{equation}
\vec{\varepsilon}=\varepsilon \left(
\begin{array}{cc}
\cos ^{2}\theta -\sigma \sin ^{2}\theta & (1+\sigma )\cos \theta \sin \theta
\\
(1+\sigma )\cos \theta \sin \theta & \sin ^{2}\theta -\sigma \cos ^{2}\theta%
\end{array}%
\right) ,  \label{EQ9}
\end{equation}%
where $\varepsilon $\ is the tensile strain, and $\sigma $\ is the Poisson's
ratio for graphene, $\sigma =0.165$. $\theta $\ is angle between the
direction of strain $\vec{\varepsilon}$ and $x$\ direction. The modification
of the atom distances distorts the reciprocal lattice as well, The primitive
vectors of the reciprocal space changes according to
\begin{eqnarray}
\mathbf{b}_{1} &\simeq &\frac{2\pi }{\sqrt{3}a}(-1+\varepsilon _{11}-\frac{%
\varepsilon _{12}}{\sqrt{3}},\frac{1}{\sqrt{3}}+\varepsilon _{12}-\frac{%
\varepsilon _{22}}{\sqrt{3}}),  \notag \\
\mathbf{b}_{2} &\simeq &\frac{2\pi }{\sqrt{3}a}(1-\varepsilon _{11}-\frac{%
\varepsilon _{12}}{\sqrt{3}},\frac{1}{\sqrt{3}}-\varepsilon _{12}-\frac{%
\varepsilon _{22}}{\sqrt{3}}),  \label{EQ10}
\end{eqnarray}

Let $\hat{\delta}$, $\hat{\delta}_{-a_{1}}$, and $\hat{\delta}_{-a_{2}}$ be
the vectors connecting NN in the honeycomb,
\begin{equation*}
t_{\hat{\delta}}=t(1-\beta \varepsilon _{22})
\end{equation*}%
\begin{equation*}
t_{\hat{\delta}_{-a_{1}}}=t[1-\beta (\frac{3}{4}\varepsilon _{11}-\frac{%
\sqrt{3}}{2}\varepsilon _{12}+\frac{1}{4}\varepsilon _{22})]
\end{equation*}%
\begin{equation}
t_{\hat{\delta}_{-a_{2}}}=t[1-\beta (\frac{3}{4}\varepsilon _{11}+\frac{%
\sqrt{3}}{2}\varepsilon _{12}+\frac{1}{4}\varepsilon _{22})],  \label{EQ11}
\end{equation}%
where $\beta \approx -\partial \log t/\partial \log a\sim 2-3.$

The Rashba spin-orbit interaction in the strained honeycomb is

\begin{equation*}
\lambda _{R_{\hat{\delta}}}=t(1+\beta ^{\prime }\varepsilon _{22})
\end{equation*}%
\begin{equation*}
\lambda _{R_{\hat{\delta}_{-a_{1}}}}=t[1+\beta ^{\prime }(\frac{3}{4}%
\varepsilon _{11}-\frac{\sqrt{3}}{2}\varepsilon _{12}+\frac{1}{4}\varepsilon
_{22})]
\end{equation*}%
\begin{equation}
\lambda _{R_{\hat{\delta}_{-a_{2}}}}=t[1+\beta ^{\prime }(\frac{3}{4}%
\varepsilon _{11}+\frac{\sqrt{3}}{2}\varepsilon _{12}+\frac{1}{4}\varepsilon
_{22})].  \label{EQ12}
\end{equation}%
Here we assume $\beta =\beta ^{\prime }$.

\section{Results and Discussion}

Graphene has been an extremely active subject of research in recent years.
The main difference between the surface of a topological insulator and that
of graphene is that the topological insulator has only one Dirac point (or
valley) and no spin degeneracy, whereas graphene has two Dirac points and is
spin degenerate.\cite{RevModPhys.81.109,moore2010birth} This difference has
far-reaching consequences, including the possibility of generating new
particles that have applications in quantum computing. Several ways have
been put forward in the literature to discuss graphene's topological
property, including breaking the symmetry of a graphene sheet by depositing
it on a substrate, controlling its band structure by doping, and adjusting
the electronic properties of a graphene bilayer by applying an external
electric field perpendicular to the surface. Graphene is also amenable to
other external influences, including mechanical deformation, which offers a
tempting prospect of controlling graphene's properties by strain.
Strain-induced pseudomagnetic fields in graphene have been explored as a
potential approach to engineering the electronic states of graphene.\cite%
{PhysRevB.79.205433,PhysRevLett.115.245501,guinea2010energy} In this work,
we will investigate the topological phase in bulk graphene and finite
nanoribbon, discussing the effect of strain.

\subsection{Bulk phase diagram}

\begin{figure}[t]
\center{\includegraphics[clip=true,width=\columnwidth]{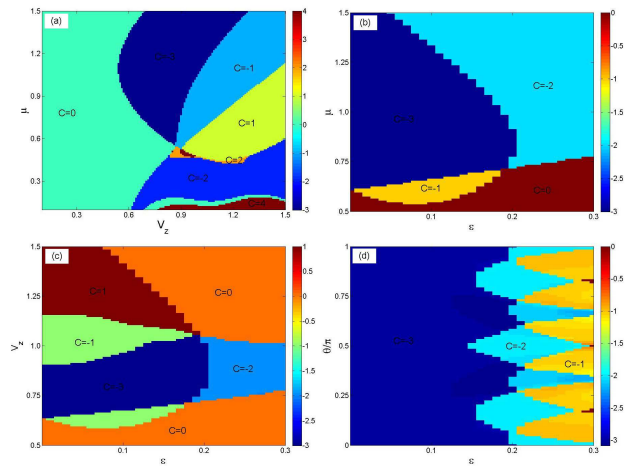}}
\caption{The phase diagram of one layer graphene sheet obtained by Chern
number. (a) Chern number as function of chemical potential $\protect\mu $
and Zeeman field $V_{Z}$ \ without strain. (b) Chern number vs $\protect\mu $
and uniform strain $\protect\varepsilon $, $V_{Z}=0.8$, $\protect\theta =0$.
(c) Chern number vs $V_{Z}$ and uniform strain $\protect\varepsilon $, $%
\protect\mu =0.8$, $\protect\theta =0$. (d)\ Chern number vs $\protect\theta
$ and uniform strain $\protect\varepsilon $, $\protect\mu =0.8$, $V_{Z}=0.8 $%
.\ Other parameters: the nearest neighbor hopping without strain $t_{0}=1$,
superconducting order $\Delta =0.5$, Rashba SOI $\protect\lambda _{R}=0.2$,
and $\protect\sigma =0.165$. }
\label{fig2}
\end{figure}

Topological orders play a crucial role in the classification of various
phases in low-dimensional systems, and the Chern number, due to the Berry
potential induced in the Brillouin zone and characterize the integer quantum
Hall effect (QHE), is one of the most typical topological invariant.\cite%
{PhysRevB.75.121403} The gap-closing condition of $H_{BDG}$ gives out
possible topological phase transition points since the topological invariant
cannot change without closing the bulk energy gap. In Fig. 2, the quantum
phase transition with Chern number of the bulk graphene electron system is
established. When the pristine graphene with SOI couples to s-wave
superconductor, this hybrid graphene system is always in insulating phase, $%
C=0$. In Fig. 2(a), the magnetic field is switched on, perpendicular to the
graphene surface, and the strong Zeeman field can induce the topological
phase. When the Zeeman field $V_{Z}\leq 0.6$, the system is always in
insulating phase $C=0$. With the increase of $V_{Z}$, some other topological
phases appear and the phase transition occurs.\ The odd Chern number
corresponds to topological nontrivial phase, where the Majorana fermion is
created, while the even Chern number corresponds to topological trivial
phase. In Fig. 2(b)-(d), the strain-induced pseudomagnetic field can sharply
change the topological property of graphene system. For low levels of
electron doping, shown in Fig. 2(b), $\mu \leq 0.55$, the system is
insulator even under strong magnetic and pseudomagnetic field. As $\mu
\simeq 0.6$, the pseudomagnetic field induced by uniform strain along zigzag
edge direction results in the topological phase transition, $C=0\rightarrow
-1\rightarrow 0$. With the increase of electron doping, the graphene
transforms into topological nontrivial phase, $C=-3$. But uniform strain
have negative influence on the stability of this nontrivial phase, where the
larger the doping, the smaller the critical phase transition strain, $%
\varepsilon _{c}$. In Fig. 2(c), $\mu =0.8$, corresponding size strain can
help to induce nontrivial phase, in particular $\varepsilon \simeq 0.1$, a
smaller magnetic field is enough to create topological nontrivial phase, $%
C=-1$. But if the Zeeman field is large enough, the critical $\varepsilon
_{c}$ also decreases, similar to topological transport behavior as function
of $\mu $ and $\varepsilon $. The main results in Fig. 2(d) present the
influence of the direction of strain. Strong uniform strain can cause the
system to be always in trivial phase when $\theta =0$, but continuous phase
transition occurs, $C=-3\rightarrow -2\rightarrow -1$, if the direction of
the strain shifts away from zigzag edge. The reason may be the effective
pseudomagnetic field along zigzag edge decreases. The obtained results
indicate that the strain can be used as a potential way to control Majorana
fermions.

\begin{figure}[t]
\center{\includegraphics[clip=true,width=\columnwidth]{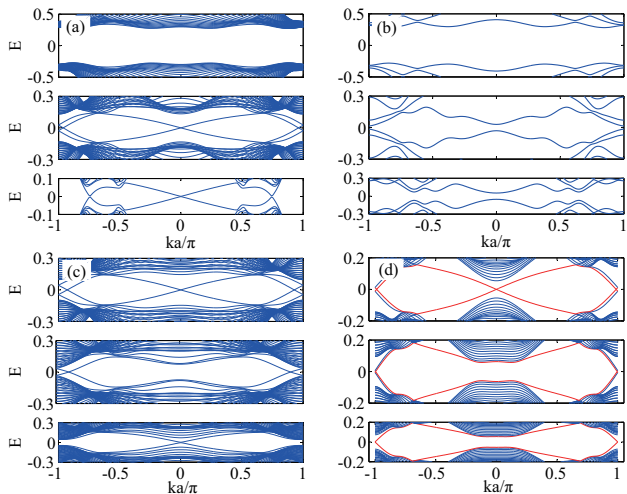}}
\caption{The band structure of infinite zigzag nanoribbon. (a) The width of
the ribbon is $W=40$, and Zeeman field $V_{Z}=0.3 (Top),0.8(Middle)$, and $%
1.2 (Bottom)$, respectively. (b) The width of the ribbon is $W=4$, and
Zeeman field $V_{Z}=0.3,0.8$, and $1.2$, respectively. Other parameters: $%
t_{0}=1$, $\protect\mu=0.8$, $\Delta =0.5$, $\protect\lambda _{R}=0.2$, $%
\protect\theta =0$, $\protect\varepsilon =0$\ and $\protect\sigma =0.165$.
(c)-(d) The band structure of different phases in the presence of strain.
(c) (Top)$\protect\mu=0.8$, $V_{Z}=0.8$, $\protect\theta =0$, $\protect%
\varepsilon =0.1$. (Middle) $\protect\mu=0.8$, $V_{Z}=0.8$, $\protect\theta %
=0$, $\protect\varepsilon =0.25$. (Bottom) $\protect\mu=0.7$, $V_{Z}=0.7$, $%
\protect\theta =0$, $\protect\varepsilon =0.1$. (d) (Top)$\protect\mu=0.8$, $%
V_{Z}=0.8$, $\protect\theta =0.5\protect\pi$, $\protect\varepsilon =0.1$.
(Middle) $\protect\mu=0.8$, $V_{Z}=0.8$, $\protect\theta =0.5\protect\pi$, $%
\protect\varepsilon =0.2$. (Bottom) $\protect\mu=0.8 $, $V_{Z}=0.8$, $%
\protect\theta =0.4\protect\pi$, $\protect\varepsilon =0.25$. }
\label{fig3}
\end{figure}

Topological quantum numbers for the bulk can often be related with those for
the edge states in finite systems.\cite{PhysRevB.74.205414} With this
bulk-edge correspondence topological properties which can be hidden in the
bulk may thus become visible around the boundaries. The emergence of chiral
edge states in the bulk gap is intimately related to the topological
property of the bulk Bloch states in the valence bands. In Fig. 3, the band
structure of infinite zigzag ribbon with width of forty unit cells ($W=40$)
in different topological phases are presented, which is characterized by
different number of crossings at zero-energy. It is found that the number of
gapless chiral edge states along the edge of the graphene ribbon equals to
the the absolute value of Chern number. When the system is insulator, $C=0$,
a well defined gap exists, while $C=-3$\ corresponds to three zero-energy
crossings as well as one crossing for $C=-1$. In Fig. 3(b), the used
parameters are the same as Fig. 3(a) except for the width of the ribbon, $%
W=4 $. It is noted that a finite gap between valence and conduction bands is
always there. The reason is that the width of the ribbon is so narrow and
the zigzag edge states can hybridize to each other. These results can also
be verified by the eigenvalue calculation of finite nanoribbon with open
boundary condition. It is also interested to note the difference of the
zero-energy crossings between $\theta =0$ and $\theta \neq 0$, shown in Fig.
3(d). When $\theta \neq 0$, the continuous phase transition occurs, $%
C=-3\rightarrow -2\rightarrow -1$, but the only one zero-energy state
crosses at $k=\pi $ for $C=-1$. In Fig. 4, the band gap of infinite graphene
zigzag nanoribbon with width of forty unit cells ($W=40$) at $k=0$ and $\pi $
is calculated, as well as the number of gapless chiral edge states. In Fig.
4 (a), (c), (e), (g), the energy gap at $k=0$\ as function of $\mu $, $V_{Z}$%
, $\theta $, and $\varepsilon $\ presents two dispersive edge states
crossing at zero energy if the system is in the topologically non-trivial
phase. Fig. 3(h) shows the energy gap at $k=\pi $ as function of strain and
its direction, it is found that the edge states crossing at $k=\pi $ as $%
\theta \neq n\pi $ ($n=0,1,2....$). In Fig. 4(b), (d), (f), (i), the number
of gapless chiral edge states is counted, corresponding to the the absolute
value of Chern number. The phase diagram obtained by Chern number is also
demonstrated by band structure calculations.

\begin{figure}[t]
\center{\includegraphics[clip=true,width=\columnwidth]{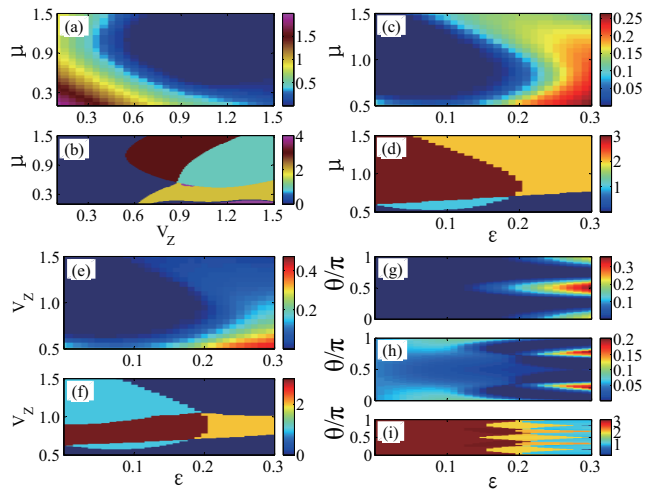}}
\caption{The edge states of infinite zigzag nanoribbon with width of 40 unit
cell ($W=40$). (a), (c), (e), (g) The energy gap at $k=0$, (h) the energy
gap at $k=\protect\pi$. (b), (d), (e), (i) The number of zero mode pairs is
shown, odd number corresponds to topological nontrivial phase, even number
is topological trivial phase and zero is insulating phase. Other parameters
are the same as Fig. 2.}
\label{fig4}
\end{figure}

\subsection{Edge states in finite graphene nanoribbon}

Designing graphene nanoribbons (GNRs) open the way to a breakthrough
carbon-based electronics, in which the lateral quantum confinement opens an
electronic gap. Recent works have shown that a combination use of
lithographic patterning of graphene samples and chemical methods such as
solution-dispersion and sonication,\cite{PhysRevLett.98.206805} the
designing GNRs can have narrow edges with controlled orientation, spatial
localization, and electronic properties.\cite{PhysRevLett.101.096402} In this
part, the investigation of the topological property and phase transition for
narrow GNRs shows that the uniform strain-induced pseudomagnetic field have
negative influence on the stability of topological nontrivial phase. In Fig.
5, a fixed size ribbon with length $L=300$\ and width $W=4$ unit cells is
chose to couple with s-wave superconductor, and its topological behavior is
similar to a quasi-1D nanowire, where the zero-energy modes localize at the
two ends of ribbon. A spinless quantum dot coupled to ribbon end is used as
scanning tunneling microscope (STM) to verify the topological phase
transition. Universal zero-bias anomalies provides a powerful method to infer
the existence of Majorana mode, in particular, it is in the topological
nontrivial phase, $G_{peak}=1/2$ $(e^{2}/h)$, in contrast to that for a dot
coupled to a regular fermionic zero mode, $G_{peak}=0$, and its
topologically trivial phase, $G_{peak}=1$.\cite%
{PhysRevB.84.201308,wang2014topological} In Fig. 5(a), the appearance of
Majorana zero mode required high level of electron doping and strong
magnetic field. Upon increasing the Zeeman field, the ribbon undergoes a
series of topological quantum phase transition as it traverses successive
topological trivial and nontrivial phases. Fig. 5(b) presents the
combination effect of electron doping and uniform strain affects the
stability of Majorana zero mode, where strong strain destroys the
topological nontrivial phase at small $\mu $, but it nearly has no influence
on the nontrivial phase stability at high $\mu $. The magnetic field and
strain-induced pseudomagnetic field would have opposing effect, shown in
Fig. 5(c). At moderate magnetic field, strong strain induces the topological
phase transition from nontrivial to trivial phase, and successive phase
transition undergoes at strong magnetic field. Fig. 5(d) indicates that the
effective pseudomagnetic field along zigzag edge changes with the direction
of the strain, and strain along armchair direction can have sharply effect,
requiring small critical strain $\varepsilon _{c}$.

\begin{figure}[t]
\center{\includegraphics[clip=true,width=\columnwidth]{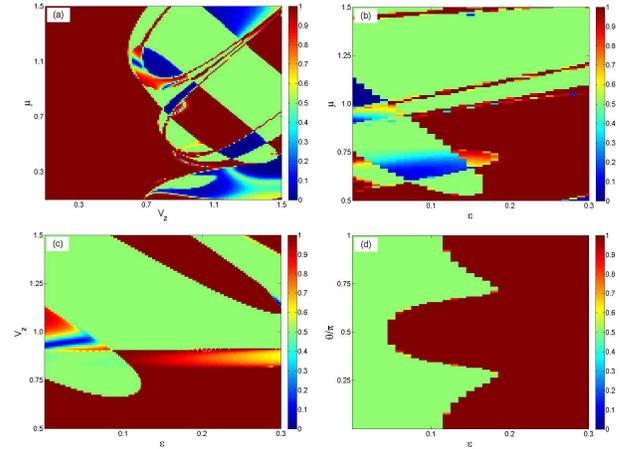}}
\caption{The pronounced zero-bias peak of detecting quantum dot is used to
verify the phase of finite nanoribbon, $A(\protect\omega )=-2\Gamma
Im(G_{1\downarrow \downarrow }^{R}(\protect\omega )+G_{2\downarrow
\downarrow }^{R}(\protect\omega ))$. Only the spin-down channel is
considered because of the large Zeeman splitting. (a) $\protect\theta =0$, $%
\protect\varepsilon =0.0$. (b) $\protect\theta =0$, $V_{Z}=0.8$. (c) $%
\protect\theta =0$, $\protect\mu =0.8$. (d) $\protect\mu =0.8$, $V_{Z}=0.8$.
Other parameters: $t_{0}=1$, $\Delta =0.5$, $\protect\lambda _{R}=0.2$, and $%
\protect\sigma =0.165$.}
\label{fig5}
\end{figure}

Recently, a local Majorana polarization is also used as order parameters to
characterize the topological transition between the trivial and nontrivial
system.\cite{PhysRevLett.108.096802} For a given eigenstate, $\Psi
^{+}=(u_{\uparrow },u_{\downarrow },\nu _{\uparrow },\nu _{\downarrow })$,
with $u$ and $\nu $ respectively the electron and hole amplitudes, the
Majorana polarizations along the $x$- and $y$-axis are defined as:%
\begin{eqnarray}
P_{M_{x}} &=&2Re[u_{\downarrow }\nu _{\downarrow }^{\ast }-u_{\uparrow }\nu
_{\uparrow }^{\ast }],  \notag \\
P_{My} &=&2Im[u_{\downarrow }\nu _{\downarrow }^{\ast }-u_{\uparrow }\nu
_{\uparrow }^{\ast }].  \label{EQ13}
\end{eqnarray}

\begin{figure}[t]
\center{\includegraphics[clip=true,width=\columnwidth]{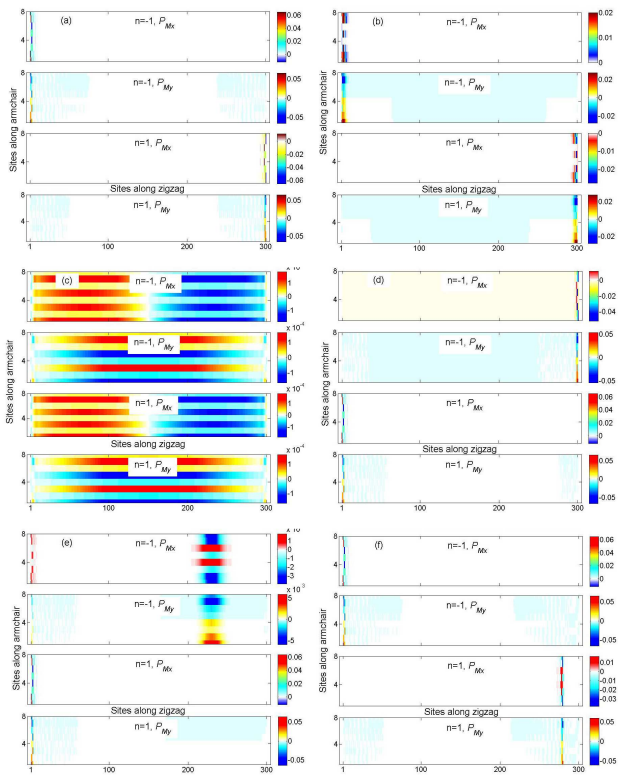}}
\caption{The Majorana polarization of two zero-modes inside the gap.
Majorana polarization is also a good order parameter to characterize the
topological transition. (a) Uniform strain along zigzag edge $\protect%
\varepsilon =0$.\ (b) $\protect\varepsilon =0.1$.\ (c) $\protect\varepsilon %
=0.25$.\ (d)\ Non-uniform strain, the strain increases as an arithmetic
series, $\protect\varepsilon _{\max }=0.05$. (e) $\protect\varepsilon _{\max
}=0.15$.\ (f) Non-uniform strain distributes as function tanh. Other
parameters: $t_{0}=1$, $\protect\mu =0.8$, $V_{Z}=0.8$, $\Delta =0.5$, $%
\protect\lambda _{R}=0.2$, $\protect\theta =0$, and $\protect\sigma =0.165$.}
\label{fig6}
\end{figure}

The Majorana polarization $P_{M_{x}}$ and $P_{My}$\ for the lowest energy
states ($n=\pm 1$) are plotted in Fig. 6. When there is no strain on the
symmetry zigzag ribbon, shown in Fig. 6(a), the system is in topological
nontrivial phase, and the values of the Majorana polarization $P_{M_{x}}$
are always opposite at the two ends of the ribbon, while $P_{My}$\ are
always equal at two ends. In Fig. 6(b), the uniform strain $\varepsilon =0.1$
along zigzag edge is switched on, the ribbon is always in its topological
nontrivial phase, and the opposite $P_{M_{x}}$ at two ends is preserved, $%
P_{My}$\ are always equal. The main difference between no strain case is
that the Majorana polarization becomes delocalized. In Fig. 6(c), $%
\varepsilon =0.25$, the ribbon is in topological trivial phase, $P_{M_{x}}$
and $P_{My}$\ are nearly zero in all sites of the ribbon. Therefore,
Majorana polarization is also a good order parameter to characterize the
topological transition. In Fig. 6(d)-(e), non-uniform strain is applied
along zigzag direction, and the strain increases as an arithmetic series
along $x$ direction, $\varepsilon _{\max }=0.05$ and $0.15$, respectively.
It is found that $P_{M_{x}}$ are always opposite but with different strength
at the two ends of the ribbon, and $P_{My}$\ are unequal.\ As $\varepsilon
_{\max }=0.15$, the topological nontrivial phase will disappear in high
strain region and topological nontrivial phase edge starts to move away from
high strain. In Fig. 6(f), the other kind of non-uniform strain is applied,%
\begin{equation}
\varepsilon =\varepsilon _{\max }\frac{1}{2}[1+\tanh (\frac{x-x_{0}}{\zeta }%
)],  \label{EQ14}
\end{equation}%
where $\varepsilon _{\max }$\ is\ maximum strain at right end, $x$\ is sites
of the ribbon, $x_{0}$\ is the sites with maximum slope, and $\zeta $\ can
control the slope of the line.\ It is also observed that the non-uniform
strain can move the Majorana polarization. For armchair graphene nanoribbon,
the values of the Majorana polarization $P_{My}$ are always opposite at the
two ends of the ribbon, and $P_{Mx}$\ are always equal at two ends. In
non-uniform graphene nanoribbon, the symmetry of the Majorana polarization
is destroyed. According to the calculation, it conforms that a MF is always
spin polarized in the sense that for a given tunnel junction, there exist a
spin-quantization axis along which it is only possible to tunnel into or out
of the state with electrons with a specific spin projection.

\begin{figure}[t]
\center{\includegraphics[clip=true,width=\columnwidth]{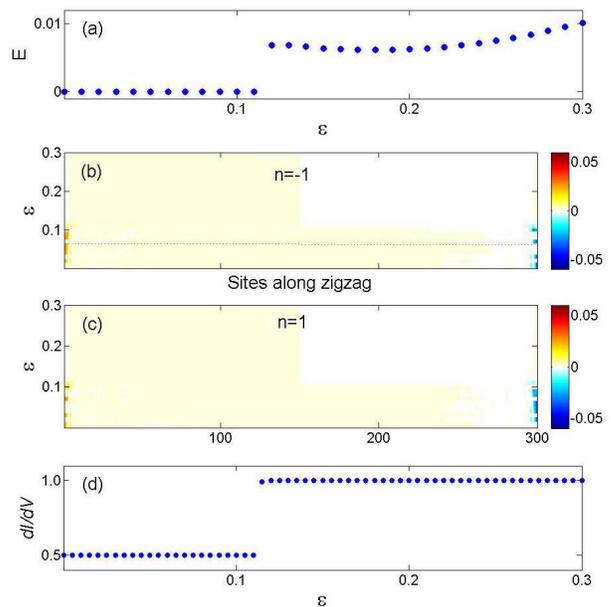}}
\caption{The topological phase transition characterized by Majorana
polarization $P_{M_{x}}$. (a) The lowest energy (n=1) as function of uniform
strain. (b)-(c) Majorana polarization $P_{M_{x}}$ vs uniform strain. (d) The
phase transition characterized by the pronounced zero-bias peak of detecting
quantum dot. $\protect\varepsilon _{c}=0.115$ is critical strain value where
topological phase transition occurs. Other parameters: $t_{0}=1$, $\protect%
\mu =0.8$, $V_{Z}=0.8$, $\Delta =0.5$, $\protect\lambda _{R}=0.2$, $\protect%
\theta =0$, and $\protect\sigma =0.165$.}
\label{fig7}
\end{figure}

In Fig. 7, the Majorana polarization $P_{M_{x}}$ and $P_{My}$\ for the
lowest energy states as function of uniform strain $\varepsilon $\ are
plotted, where $\mu =V_{z}=0.8$. Fig. 7(a) gives the energy values for
different strain indicating the phase transition point at the uniform strain
$\varepsilon \geqslant 0.115$. This phase transition behavior can also be
observed from Majorana polarization, shown in Fig. 7(b)-(c), the values of
the Majorana polarization $P_{M_{x}}$ are always opposite at the two ends of
the ribbon as the ribbon is in its topological nontrivial phase, while $%
P_{My}$\ are always equal at two ends. When the system is in trivial phase,
there is no Majorana polarization. In Fig. 7(d), the pronounced zero-bias
peak of coupled spinless quantum dot also gives the critical strain value $%
\varepsilon _{c}=0.115$ where topological phase transition occurs.
Therefore, for narrow GNRs, the uniform strain can also be applied to
control the creation and transfer of Majorana zero modes.

\begin{figure}[t]
\center{\includegraphics[clip=true,width=\columnwidth]{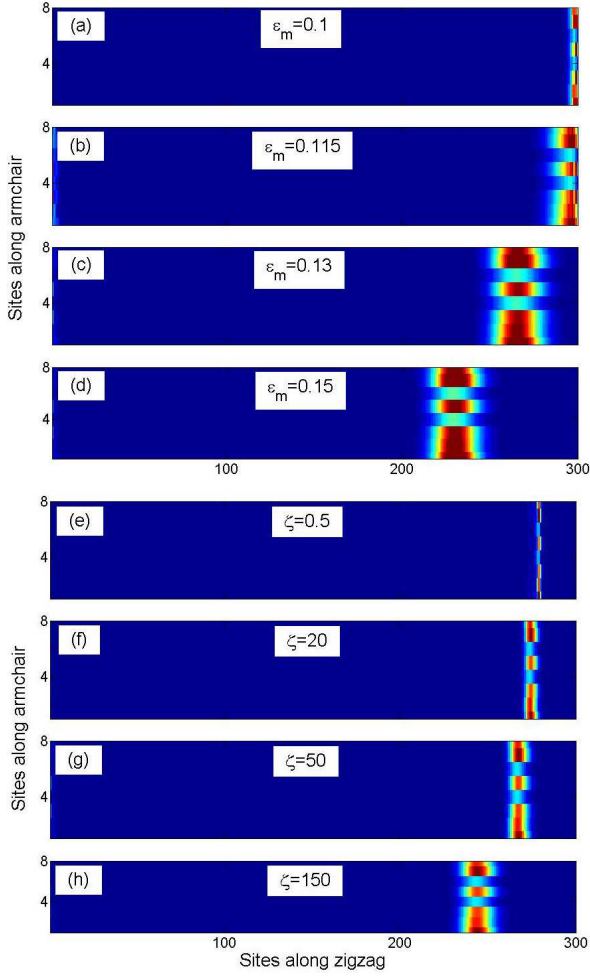}}
\caption{The critical value of strain. (a)-(d) The strain increases as an
arithmetic series. (a) $\protect\varepsilon _{\max }=0.1$. (b) $\protect%
\varepsilon _{\max }=0.115$. (c) $\protect\varepsilon _{\max }=0.13$. (d) $%
\protect\varepsilon _{\max }=0.15$. (e)-(h) The non-uniform strain is
distributed as tanh function, and $\protect\varepsilon _{c}=0.115$. If the
strain of one region exceeds $\protect\varepsilon _{c}$, the region will be
into non-topological region. (e) $\protect\zeta =0.5$, (f) $\protect\zeta %
=20 $, (g) $\protect\zeta =50$, and (h) $\protect\zeta =150$. Other
parameters: $t_{0}=1$, $\protect\mu =0.8$, $V_{Z}=0.8$, $\Delta =0.5$, $%
\protect\lambda _{R}=0.2$, $\protect\theta =0$, and $\protect\sigma =0.165$.}
\label{fig8}
\end{figure}

Graphene nanoribbon exhibits band gap modulation when subjected to strain.
Besides uniform strain, nanoengineered non-uniform strain distribution in
graphene is aslo a promising road to generate a band gap and a large
pseudomagnetic field.\cite{PhysRevB.86.041405,tomori2011introducing}
Mechanical deformations such as nanoribbon twist, bending and inhomogeneous
deformation can lead to non-uniform strain.\cite{dobrinsky2011electronic}
Recent experiments showed that non-uniform strain can also be produced by
depositing graphene over pillars. Here, we consider two types of non-uniform
strain with different distribution rule, linear and nonlinear. Firstly, we
consider the case that the strain increases as an arithmetic series along $x$
direction, and the maximum strain localizes at right end of the ribbon. The
main results are present in Fig. 8. At the left end of the ribbon, the
strain is small and the left MF always localizes at it, but the right end
has large strain and the localization of right MF is changed. So we only
plot the wavefunction of right MF. In Fig. 8(a), as the $\varepsilon _{\max
}=0.1$,\ the wavefunction of right MF becomes delocalized. When the maximum
strain $\varepsilon _{\max }$ is large than $0.115$, shown in Fig. 8(b)-(d),
the wavefunction becomes\ more delocalization and move to low strain region.
The obtained critical strain value $\varepsilon _{c}$ of phase transition is
the same as uniform strain. Therefore, both the uniform and non-uniform
strain can provide a new way to manipulate Majorana fermions.

To get a deep insight into the non-uniform strain-induced pseudomagnetic
field, the strain is distributed nonlinear, as in Eq. (14). When $\zeta \ll
Na$, the strain will have a sharp increase at $x=x_{0}$, and becomes smooth
on other sites. As can be seen from Fig. 8(e)-(h), $\varepsilon _{\max }=0.3$
and $x_{0}=280$, the smaller the $\zeta $, the larger the slope at $x_{0}$.
When $\zeta =0.5$, the right MF localizes at $x=279$, and the high strain
region becomes non topological region, this behavior indicates that the
pseudomagnetic field induced by large strain is harmful to the stability of
topological non-trivial phase. In Fig. 8(f)-(h), $\zeta =20,50,150$,
respectively, the strain starts to spread over the whole ribbon, and the
right MF transfers to left end, localizing at $x=275,268,244$. The critical
strain is $\varepsilon _{c}=0.115$, where the right MF localizes. At the
same time, the right MF becomes delocalized. As the distance between two MFs
is much small, the wavefunction of the two MFs become overlap, and the
system transfers to topological trivial phase. When the maximum strain $%
\varepsilon _{\max }\leq $\ $\varepsilon _{c}$, the two MF will always
localize at the two ends of the ribbon, and the whole ribbon is in the
topological nontrivial phase.

\begin{figure}[t]
\center{\includegraphics[clip=true,width=\columnwidth]{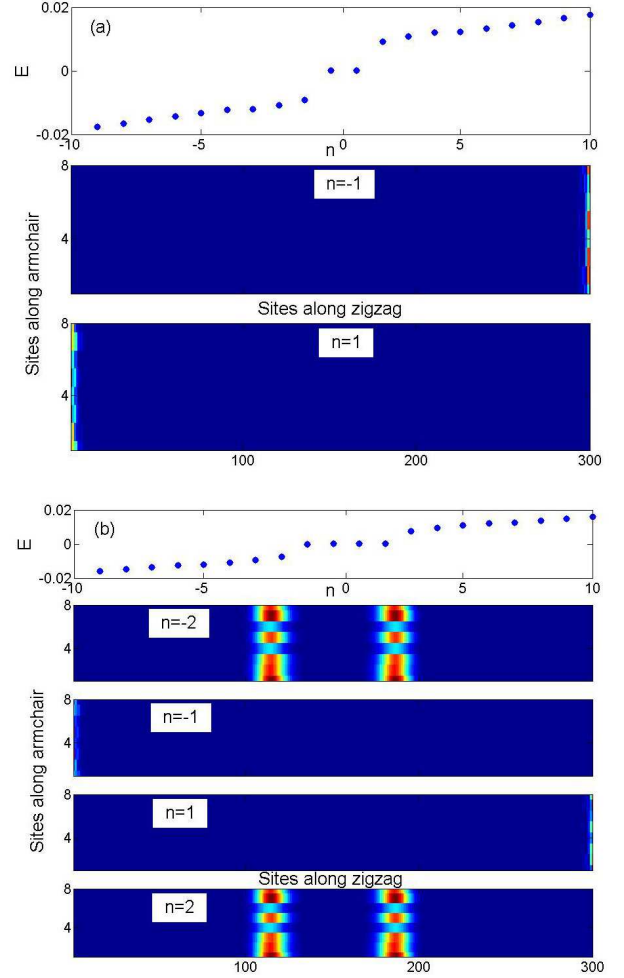}}
\caption{Non-uniform strain creates new Majorana fermions. The maximum
strain localizes at the middle of the ribbon. (a) $\protect\varepsilon %
_{\max }=0.1$. (b) $\protect\varepsilon _{\max }=0.15$. The other parameters
are the same as Fig. 8.}
\label{fig9}
\end{figure}

Furthermore, the strain distribution can be non-monotone variation, such as
bending ribbon. It is therefore interesting to study the case that the
strain linearly increases at first and then decreases to zero. In Fig. 9(a),
the maximum strain at the middle of the ribbon is $\varepsilon _{\max
}=0.1<\varepsilon _{c}=0.115$, the whole ribbon is always in its topological
nontrivial phase. But as $\varepsilon _{\max }=0.15>\varepsilon _{c}$, there
is a new window opens in the middle region of the ribbon, $\varepsilon
>\varepsilon _{c}$,\ which is trivial phase. Meanwhile, there are two
topological nontrivial regions in the ribbon, separating by a domain wall
induced by the maximum strain at the middle of the ribbon. For this reason,
there are two pairs zero-energy modes inside the gap, and the non-uniform
strain creates a new pair of Majorana fermions instead of local gate.

\subsection{Experimental realization}

\begin{figure}[t]
\center{\includegraphics[clip=true,width=\columnwidth]{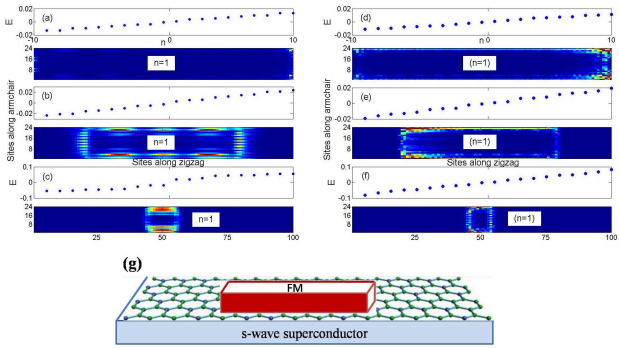}}
\caption{The Majorana fermions in the wide ribbon. Non-uniform chemical
potential helps to move topological region. The width is $W=12$ unit cell,
the length is 100 unit cell.\ (a) $l=100$. (b) $l=60$. (c) $l=10$. Other
parameters: $t_{0}=1$, $V_{Z}=0.7$, $\Delta =0.5$, $\protect\lambda _{R}=0.2$%
, $\protect\theta =0$, $\protect\varepsilon =0$\ and $\protect\sigma =0.165$%
. }
\label{fig10}
\end{figure}
\

As discussed above, designing graphene nanoribbons have been put forward
theoretically and experimentally. Next, we propose some alternative devices
without strain to form the nanoribbon boundaries in graphene sheet by use of
quantum control techniques. Here we choose a finite ribbon, with width $W=12$
and length $L=100$ unit cells, as the target systems. In Fig. 10(a)-(c),
using the local gate, the non-uniform chemical potential is distributed as
\begin{equation}
u=u_{c}-\delta \lbrack 1-\tanh (\frac{x-(x_{0}-l/2)}{\zeta })\tanh (\frac{%
-x+(x_{0}-l/2)}{\zeta })],  \label{EQ15}
\end{equation}%
where $u_{c}$\ is\ critical value to achieve topological nontrivial phase, $%
x $\ is sites of the ribbon, $x_{0}$\ is the sites with maximum slope, $%
\zeta $\ can control the slope of the line, $\delta $ is the gap away from $%
u_{c}$, $l$\ is the length of topological nontrivial region. When the ribbon
has large length and width, but length is much large than width, the system
looks like a infinite nanoribbon and the Majorana zero-energy states
localize at the edges of the ribbon. If the width is very narrow, the ribbon
is similar to a nanowire and the Majorana end states localize at the two end
of the ribbon. Fig. 10 gives comparable width to the length of finite
ribbon, and the zero-energy states localize around the four edges, forming a
rectangle region. From Fig. 10(a)-(c), the rectangle's area decreases with $%
l $, suggesting the non-uniform potential defined topological nontrivial
region. In Fig. 10(d)-(f), non-uniform Zeeman field is introduced to
control the topological nontrivial area, depending on the size of
ferromagnetic, as the sketch in Fig. 10(g). The magnetic field is written as
\begin{equation}
V_{z}=\frac{(V_{z})_{c}}{2}[1-\tanh (\frac{x-(x_{0}-l/2)}{\zeta })\tanh (%
\frac{-x+(x_{0}-l/2)}{\zeta })],  \label{EQ16}
\end{equation}%
where $(V_{z})_{c}$\ is\ critical value to achieve topological nontrivial
phase. Similar results are obtained to non-uniform chemical potential.

\section{Conclusion}
In this work, the topological phase transition in bulk graphene and
nanoribbon are discussed in detail by calculating the Chern number, band
structure and wavefunction. The Majorana fermions in bulk graphene and
finite nanoribbon localize at different position, where the topological
property hidden in the bulk becomes visible around boundaries, and MF
wavefunction of narrow ribbon localizes at the two ends. According to the
calculation of strain-induced pseudomagnetic field, it is observed that
strain has negative influence on the stability of topological nontrivial
phase either uniform or nonuniform. This special property provides a new way
to transfer, create and fuse Majorana fermions, which is useful in designing
topological quantum computation. For experimental realization, the
nonuniform chemical potential, Zeeman field and strain can be used as
quantum control techniques to form ribbon of arbitrary shape, tailoring
functionalized graphene.

\section*{Acknowledgement}

\end{document}